\documentclass{IEEEtran}
\IEEEoverridecommandlockouts
\usepackage{cite}
\usepackage{url}  
\usepackage{lscape}
\usepackage{hologo}
\usepackage{amsmath,amssymb,amsfonts}
\usepackage{algorithmic}
\usepackage{graphicx}
\usepackage{textcomp}
\usepackage{xcolor}
\usepackage{flushend}
\def\BibTeX{{\rm B\kern-.05em{\sc i\kern-.025em b}\kern-.08em
    T\kern-.1667em\lower.7ex\hbox{E}\kern-.125emX}}
\begin{document}

\title{Assessing the Impact of AR-Assisted Warnings on Roadway Workers' Stress \\Under Different Workload Conditions\\
}



\author{
    \vspace{25 pt}
    \begin{minipage}{0.3\textwidth}
        \centering
        \IEEEauthorblockN{Fatemeh Banani Ardecani}
        \IEEEauthorblockA{
            \textit{University of North Carolina at Charlotte}\\
            USA \\
            fbanania@charlotte.edu}
    \end{minipage}%
    \hfill
    \begin{minipage}{0.3\textwidth}
        \centering
        \IEEEauthorblockN{Amit Kumar}
        \vfill
        \IEEEauthorblockA{
            \textit{University of North Carolina at Charlotte}\\
            USA \\
            akumar28@charlotte.edu}
    \end{minipage}%
    \hfill
    \begin{minipage}{0.3\textwidth}
        \centering
        \IEEEauthorblockN{Omidreza Shoghli}
        \vfill
        \IEEEauthorblockA{
            \textit{University of North Carolina at Charlotte}\\
            USA \\
            oshoghli@charlotte.edu* \thanks{*Corresponding Author}}
    \end{minipage}
}
\maketitle

\begin{abstract}
Recent data from the Federal Highway Administration highlights an alarming increase in fatalities and injuries in roadway work zones, emphasizing the need for enhanced worker safety measures. This study addresses this concern by evaluating stress levels among roadway workers equipped with AR-assisted multi-sensory warning technology during varying work intensities. The research leverages a high-fidelity Virtual Reality environment to simulate realistic work scenarios, enabling safe evaluation of high-risk situations.  Unlike previous studies focusing on external factors, this research investigates the internal physiological impact on workers. Utilizing wearable sensors, the study collected physiological data, including photoplethysmography (PPG), electrodermal activity (EDA), and skin temperature (ST), to assess stress levels continuously and non-invasively. Our findings from 18 participants reveal significant differences between light- and medium-intensity activities in heart rate variability metrics. These metrics commonly used to assess autonomic nervous system function and stress levels, included mean heart rate, NN50, pNN50, and HF-HRV. By examining the relationship between AR-enabled warnings, work intensity, and stress levels, the study contributes to enhancing worker safety and well-being. The proposed methodology offers potential for active stress monitoring in the field, contributing to enhanced safety practices and worker productivity in construction sites. By providing real-time physiological data, this approach enables informed stress management and more effective hazard warning systems in roadway work zones. This research bridges a gap in understanding the physiological impacts of AR-assisted warnings on roadway workers. The insights gained from this study can inform future safety interventions and guide the development of more effective warning systems.

\end{abstract}

\begin{IEEEkeywords}
Worker's Stress; Roadway workers; Work zone safety; Multi-sensory AR warnings; Virtual Reality
\end{IEEEkeywords}

\section{Introduction}

Fatal traffic crashes at roadway work zones have been on the rise in recent years. According to the Federal Highway Administration (FHWA) work zone facts and statistics report, there was a significant increase in fatal crashes at work zones from 863 in 2020 to 956 in 2021 \cite{FHWA}. Research has shown that excessive stress in workers heightens the risk of errors, injuries, and various health issues among workers while concurrently being associated with reduced productivity levels \cite{jebelli2018eeg}.   
In roadway work zones, lane closures, proximity to moving vehicles, night shifts, and the presence of construction vehicles and employees can all cause stress in workers. A construction worker's behavior can be significantly affected by stress \cite{lim2017analyzing}. The hormones released by the brain under stress, including cortisol and adrenaline, may have an impact on cognitive processes, judgment, and response times.

In the wake of rising highway worker fatalities at road construction sites, there's an immediate need to create advanced safety systems that safeguard workers. Augmented Reality (AR) shows promising potential in alerting workers, but its implementation in road work zones and its impact on stress and attention have not been thoroughly investigated. Currently, most of the research is focused on external factors like lane closures, traffic congestion, warnings, determining safe work radius, and traffic management. While this is important, the body of knowledge is missing an integrated effort to analyze how these factors affect the worker's stress. Therefore, this study aims to assess the effects of two categories of work (light- or medium-intensity activity) on stress levels experienced by roadway workers when they were receiving multi-sensory warnings during their routine tasks.

This study aims to assess the stress levels experienced by roadway workers as they receive multi-sensory AR-enabled warnings during their routine tasks, examining the effects within two categories of work as either light- or medium-intensity activity. Furthermore, it leverages a high-fidelity virtual reality environment for safe evaluation and testing of rare high-risk scenarios. Additionally, it introduces a model that enables continuous and non-invasive monitoring of stress levels among roadway workers based on work activity. The proposed methodology can be leveraged for active monitoring of stress in the field, thereby enhancing safety at construction sites and promoting the well-being and productivity of workers.

\section{Literature Review}
\label{sec:Preparation}
\subsection{Mental Stress in Construction Workers}
The number of mortality and disability cases involving construction workers was the highest amongst the major industrial sectors \cite{pan2021roles}. 
Current approaches investigating the physical demands of various tasks provide valuable information for evaluating certain construction activities. However, these approaches often focus on specific individual traits, such as physiological characteristics, and environmental factors, such as ambient temperature and humidity. In simpler terms, this means that employees may exert varying levels of effort when performing the same task due to these individual and environmental differences \cite{jebelli2019application}. 
In individuals working under continuously demanding and stressful conditions, emotional stress manifests in chronic fatigue, emotional drain, and a loss of devotion to job duties. Stress can lead to reduced attention on work tasks and subsequently cause one to ignore safety behaviors, thereby increasing injury incident rates \cite{leung2012preventing}.

Molen et al.\cite{van2000work} presented a qualitative study on the nature and feasibility of measures to reduce work stress. Construction work conditions can induce anxiety that may be high enough to elicit a robust fear response. Fear and anxiety are known to have different neuro-anatomical substrates and physiological outcomes and may lead to differences in behavior and action tendencies \cite{davis2009relationship}.

Because they are at the bottom of the organizational hierarchy, construction workers have little authority over their work. 
These organizational stressors not only stress out the construction workers but also affect how they behave in terms of safety. The primary factor behind most construction workers' injury events is a deterioration in safety behaviors \cite{leung2012preventing}. In the case of roadway construction and maintenance, dynamic worksite conditions and proximity to live traffic frequently expose highway workers to unsafe work zone proximity, resulting in accidents \cite{sabeti2023mixed, sabeti2021toward, sabeti2022toward}. 

\subsection{Stress Monitoring Using Wearable Sensors}
Everyone reacts to stress differently; thus, measuring and tracking stress can be difficult. Psychoanalysis, human sensing, and medical examinations could be used to identify the signs and symptoms of stress. Basic signs of elevated stress include headaches, stiff muscles, sleeplessness, and rapid heartbeat \cite{hou2015eeg}.
In the clinical setting, physiological indicators of biochemical reactions, such as stress hormones, have been widely used as accurate stress markers. Because stress-related hormones, such as cortisol and glucocorticoids, alter in response to stress, monitoring these hormone changes can give us useful information about how stressed people are \cite{ranabir2011stress}. Though this strategy is effective and desirable, it is impractical for continuous stress monitoring at an active construction site because serum, saliva, urine, or hair samples must be taken repeatedly to measure stress-related hormones. Continuous monitoring of stress is particularly important on construction sites due to the constantly changing nature and challenges at the workplace. Additionally, the analysis of the collected biological samples necessitates laboratory processing, which is challenging to implement in the field. To bypass this issue, recent studies \cite{arabshahi2021review, nnaji2021wearable,sabeti2022wearable} have utilized wearable sensors for the assessment of stress levels. 
For this purpose, physiological signals, which are generated by the body's processes, can be collected. These signals have traditionally been categorized into two groups. The first group includes electrical physiological signals such as electrodermal activity (EDA) \cite{shakerian2021assessing}, electroencephalography (EEG) \cite{qin2023electroencephalogram}, heart rate (HR), heart rate variability (HRV), electrocardiography ECG \cite{rostamzadeh2023stress}, photoplethysmography (PPG) \cite{lee2021assessment}, peripheral skin temperature (ST) \cite{umer2022simultaneous}. The second group comprises non-electrical physiological signals, like inertial measurement units (IMU) and potential of hydrogen (pH). Variations in electrical physiological signals are closely linked to stress levels \cite{jebelli2019application}. Accurately determining workers' stress levels in the field can lead to early recognition of stress, thereby enhancing both safety and productivity at construction sites.

A smart sensor in a wristband can provide the signal known as EDA. By measuring the fluctuation in skin conductance, EDA provides information about the electrical characteristics of the skin.  An EDA sensor inserts a low, steady voltage into a wristband-style sensor. Next, the voltage variations brought on by sweat gland activity are monitored. A wristband-style sensor contains an infrared thermopile that measures the ST signal. An infrared thermopile uses the skin's infrared energy to measure temperature. A higher temperature is correlated with higher infrared energy. \cite{betti2017evaluation, gjoreski2017monitoring}. 
Heart signal analysis allows for the computation of heart activity characteristics known as heart rate variability (HRV). These characteristics result from processing the heart signal across time, frequency, and nonlinear domains. For instance, in the time domain, an HRV metric is the root mean square of successive heartbeat intervals (R-R intervals), termed RMSSD. Frequency domain HRV metrics include the power within specific frequency bands (such as high frequency - 0.15 to 0.4 Hz and low frequency - 0.04 to 0.15 Hz) denoted as HF and LF within the HRV spectrum. Lastly, in the nonlinear domain, an HRV feature example is the entropy calculated from beat-to-beat intervals.

Earlier research has demonstrated that these characteristics are associated with specific human conditions. For instance, stress levels have been linked to a reduction in RMSSD, while an increase in HF (high frequency) is associated with cognitive load \cite{lohani2019review}. Electrodermal activity (EDA), also known as Galvanic Skin Response (GSR), has similarly exhibited correlations with human conditions like stress and workload. EDA signals are typically broken down into two primary components: tonic and phasic. Tonic represents long-term changes in the signal, while phasic accounts for momentary shifts in the EDA signal. The tonic aspect helps establish the skin conductance level (SCL), while the phasic component defines the skin conductance responses (SCR). Both SCL and SCR have been found to be linked with heightened cognitive load and stress levels.

\section{Method}
To achieve the objective of the study, we designed an experiment in a high-fidelity Virtual Reality environment replicating a real-world roadway work zone. The study protocol (21-0357) was reviewed and approved by the Institutional Review Board (IRB) at the University of North Carolina at Charlotte. 
Prior to data collection, all individuals were briefed about the data-collection procedure, and had the option of choosing to opt out of participation at any time without giving any explanation.  
None of the participants mentioned any physiological or physical issues impacting their ability to function at work. The participants were asked to perform two routine roadway maintenance tasks in the VR environment while wearing the sensing technologies and warning delivery devices. Three members of the research team recorded the activities over a half-hour session. 

\begin{figure}[!htb]
    \centering
    \includegraphics[width=0.5\textwidth]{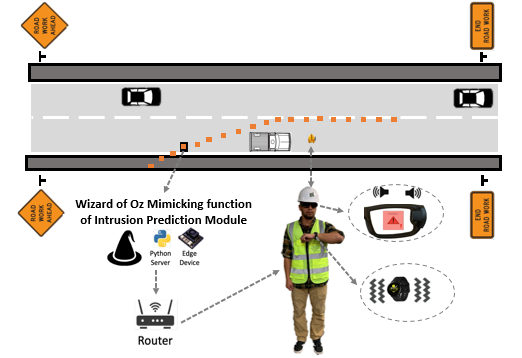}
    \caption{Outline of the AR-assisted Warning System}
    \label{fig_1}
\end{figure}

AR-assisted safety protocols within roadwork zones combined with warnings can extensively impact the response to danger and safety of workers as shown in Figure \ref{fig_1} Thus, it is indispensable to determine the worker's response to stress in a roadway work zone to gain a better understanding of the usability and effectiveness of the AR-assisted safety technology.

In this study, we evaluated the metrics related to the skin and cardiovascular system. To achieve this, we employed a wristband-type wearable equipped with embedded sensors, including photoplethysmography (PPG), skin temperature (ST), and electrodermal activity (EDA). The PPG sensor analyzes the difference between transmitted and reflected light and quantifies the variations in blood pulse within the arteries. This difference served as an indicator of heart activity. Additionally, we investigated cardiovascular parameters such as heart rate (HR) and heart rate variability (HRV). These parameters were derived by further analysis of the obtained data from the PPG sensor. It should be noted that PPG technology is commonly integrated into smart wearables and can be effectively utilized in both naturalistic and experimental settings.

\subsection{Experiment Design}

The experiment design consisted of participants triggered by warnings at certain time intervals while carrying out maintenance activities categorized into light and medium, as follows:  

\textit{Warning Trigger}: This consisted of three warning stimuli that were administered to the participants while they carried out the task in the VR environment. All three multi-sensory warnings were intended to communicate predicted risk to the workers by delivering haptic, audio, and visual cues at the same time. 
Tizen Native framework was used to administer the haptic stimulus, which utilizes a predefined pattern available in the API \cite{tizen}. The audio warning was a high-pitched beep sound with a frequency of 44,100 Hz and a duration of 0.2 milliseconds. The warnings were designed to function simultaneously, without any delay, as soon as the back-end system activated them. Moreover, the visual warnings about potential hazards were delivered using AR simulation in the VR environment. 

\textit{Activity Types}: The construction tasks were categorized as low and medium-intensity activities. Similar separation of tasks has been done previously based on energy-expenditure prediction program (EEEP) \cite{jebelli2019application} and \cite{virokannas1996thermal}. 

\textit{Light Activity}: These work scenarios include tasks like standing, briefing, and other tasks involving little movement of the body parts. Assembly, reading the construction drawings, and inventory work are a few examples of typical low-intensity tasks. For the purpose of low-intensity activity, an inspection task was chosen for this study, which included taking a picture of a clogged stormwater inlet by the roadway shoulder. Such incidents are a common maintenance task at roadway work zones where curbs are often blocked by overgrowth of vegetation, small debris, and fallen leaves. 

\textit{Moderate Activity}: These tasks include tasks like cleaning up the site, locating tools, moving light materials, and measuring and cutting sheetrock. For the purpose of the study, cleaning the clogged inlet drainage by using a leaf blower was selected as a medium-intensity task. Cleaning jobs, as such, are standard upkeep at roadway work zones.

\subsection{Apparatus}
\begin{figure}[!hb]
    \centering
    \includegraphics[width=0.5\textwidth]{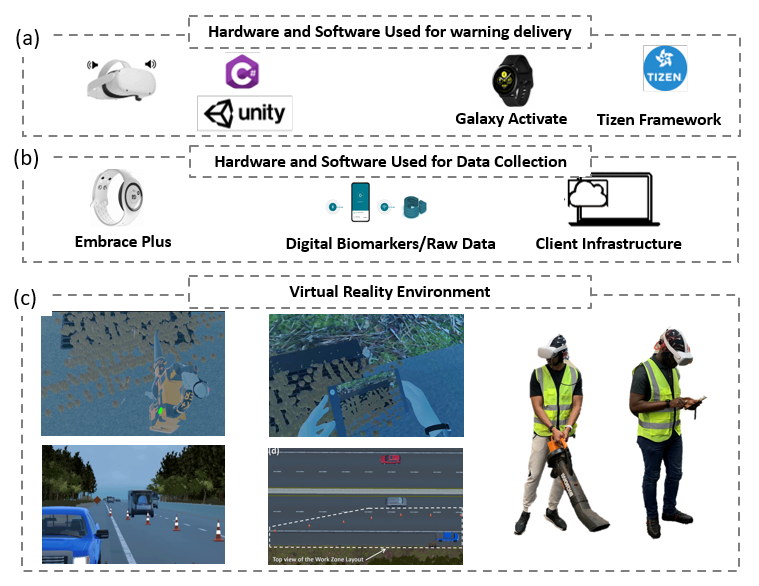}
    \caption{Apparatus of the Experiments (a) Hardware and Software Used for warning delivery, (b) Hardware and Software Used for Data Collection, and (c) Virtual Reality Environment }
    \label{fig_2}
\end{figure}

\textit{Roadway Work Zone in Virtual Reality.}
The study used the guidelines provided in the Manual on Uniform Traffic Control Devices (MUTCD) \cite{mutcd2006manual} as a reference to create a virtual highway work zone environment. Figure \ref{fig_2}(c) shows the simulated environment, including live traffic and highly detailed 3D models, closely resembling the real roadway work zone environment. 
Using the VR environment developed by the research team \cite{sabeti2024augmented}, participants were equipped with a VR headset while holding a real-life tablet (light activity) or a leaf blower (medium activity). Also, they wore a smartwatch for delivering the haptic warning and a wristband for collecting physiological data. The figure also depicts the immersive and interactive VR environment captured from the participants' viewpoints while they performed the tasks of the experiment.

The modeling utilized the Oculus Quest 2 VR headset in the Virtual Reality simulation as shown in Figure \ref{fig_2}(a) Furthermore, the Unity 3D game engine was used to develop the stream VR software. To allow the exact simulation interaction and user experience in the VR interface, the participants were able to simultaneously observe the gadgets and equipment (smartwatch, tablet, and leaf blower) in the VR while wearing the smartwatch and handling the tablet/leaf blower during the actual participation. 

\textit{Wristband Sensing Device to Capture Physiological Data.} To measure workers' physiological data in the roadway work zone environment in VR,  a wristband-type sensing technology was used. 
An off-the-shelf wearable sensing wristband, Embrace Plus, was used to collect workers' PPG, EDA, and ST signals \cite{empatica}. These were recorded simultaneously at the highest recording rate (PPG at 64 Hz and EDA and ST at 4 Hz). 

\textit{Leaf Blower, Tablet, Warning Watch, and Camera for Video recording}. 
Participants wore a Samsung Galaxy Watch, as shown in Figure \ref{fig_2}(a), that was used to deliver haptic warnings. The warnings were administered using the Tizen Native framework. Additionally, the participants used an Amazon Fire Tablet, as shown in Figure \ref{fig_2}(c), to take pictures of a clogged storm water inlet (light level activity) and a WORX WG509 12Amp leaf blower to clear the leaves (medium level activity).

\subsection{Experiment Procedure}
The experiment involved a carefully structured sequence of activities to ensure a comprehensive and ethical approach to the research. 
First, the nature of the study, its goals, and the participants' roles were explained to the volunteers. Following the consent process, administrators of the experiment explained the objectives and ensured that the participants clearly understood what was expected of them throughout the study. The experiment was designed to be completed within a reasonable time frame, with a maximum duration of up to half an hour per session. 

The participants were tasked to replicate real-world scenarios commonly encountered in roadway work zones. The focus was on taking pictures of the clogged inlet and removing obstructions from the same. To design the study, existing literature \cite{snyder2022aperture, tizen} that discusses the influence of physical activity, intensity, and cognitive load on worker's stress was utilized. 
Based on this knowledge, an obstruction removal task requiring participants to engage in higher levels of physical exertion than routine inspection of roadways was developed. 
The task began with participants capturing pictures of the clogged drain and then, as a subsequent task, activating the leaf blower and directing it towards the obstructed drop inlet within the virtual environment. As they did so, the virtual reality environment featured a carefully designed photograph capture with sound on the tablet. On the subsequent task, the blowing effect with a leaf blower sound effectively cleared the leaves positioned on top of the drop inlet. This dynamic and interactive task continued until all necessary warnings were delivered, and the administrator signaled the completion of the task.

The developed virtual work zone is depicted in Figure \ref{fig_2}(c). To enhance the realism and interactivity of the study, the model uses a mixed-reality interaction for this experiment. This approach allows the participants to simultaneously engage with physical objects in the real world and virtual objects within the virtual environment.

\subsection{Participants}

The study utilized data from 18 out of the 22 participants. The removal of the data for 4 participants was due to a technical challenge in recording and synchronizing physiological data from the study. Out of the final count (N=18), the average age of the participants was 28.27 years.
Five participants did not have work experience, while the remaining  13 had an average work experience of 3.92 years (SD =4.78). The average duration for all the participants to complete the medium and light-intensity tasks was 1 minute 42 seconds and 1 minute 59 seconds, respectively.

\subsection{Physiological Data Analysis}

\begin{figure}[!htb]
    \centering
    \includegraphics[width=0.48\textwidth]{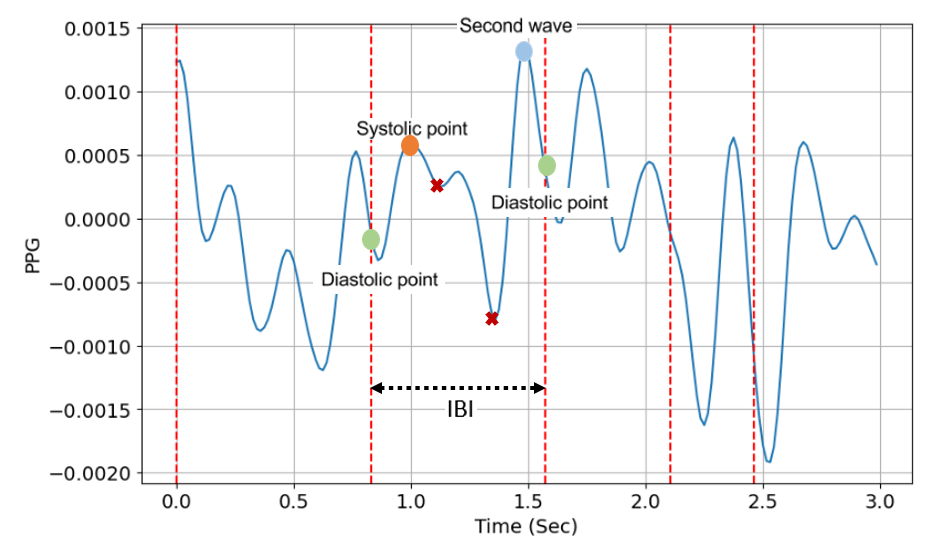}
    \caption{Interbeat Interval (IBI) defined as the time interval between two fiducial points on the diastolic pulse wave for the BVP signal}
    \label{fig_6}
\end{figure}

The Heart Rate Variability (HRV) features were computed from the interbeat interval data (IBI) as depicted in Figure \ref{fig_6} captured through the Empatica Embrace Plus device \cite{empatica}. The features span across different time, frequency, and nonlinear domains. Using the pyHRV \cite{gomes2019pyhrv} package in Python, features such as HF and RMSSD were calculated across these domains for each participant during the study. It is important to note that certain computed features, like LF, might not be suitable for short-term data collection, prompting a focus only on features applicable to shorter timeframes. Additionally, for Electrodermal Activity (EDA) analysis, denoising of EDA signals from the Empatica Embrace Plus involved using a Butterworth filter with a low cutoff frequency of 1.5 Hz \cite{jebelli2019application}.

Furthermore, a Hampel filter was utilized to eliminate outliers in the physiological data, following Allen's method \cite{ALLEN2009303}, which replaces spikes with the median value of neighboring signals. Subsequently, the modified signal was processed using the Neurokit 2 package \cite{makowski2021neurokit2}. This package allows for signal decomposition into tonic and phasic components and facilitates the computation of various Electrodermal Activity (EDA) features, including the count of phasic skin conductance responses and the skin conductance level (SCL). The study then compared the number of SCR peaks among different participants and conditions.

A paired t-test was used to compare heavy and light activity across all participants. Additionally, parameters like mean-NNI, mean-HR, std-HR, SDNN, RMSSD, NN50, PNN50, and the number of SCR Peaks were analyzed to draw inferences and conclusions from the model. RMSSD, calculated as the square root of the mean squared differences between successive NN intervals, stands as one of the most frequently utilized measures in the time domain. NN50 signifies the count of interval differences greater than 50 milliseconds between successive NN intervals, while pNN50 is the proportion calculated by dividing NN50 by the total count of NN intervals. Additionally, the Blood Volume Pulse (BVP) captured by the PPG sensor stands as a primary output. This BVP data shares a strong correlation with Interbeat Interval (IBI) data, which represents the time lapse between individual heartbeats. It is to be noted that the IBI information stems from processed PPG/BVP signals from the Embrace Plus wristband, utilizing an algorithm that effectively eliminates erroneous peaks caused by noise in the BVP signal.

\section{Result and Discussion}

\begin{figure}[!htb]
    \centering
    \includegraphics[width=0.48\textwidth]{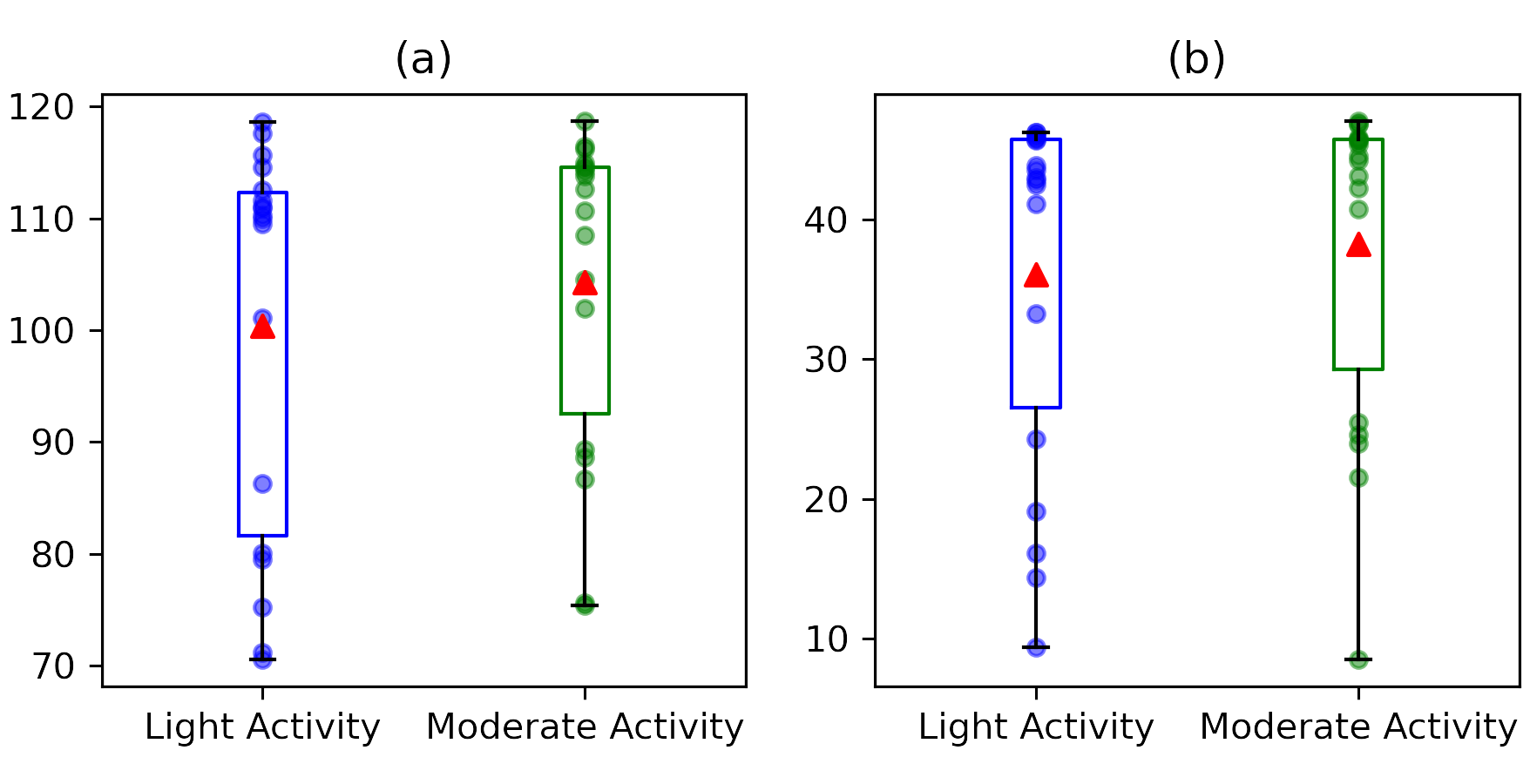}
    \caption{(a) Mean Heart Rate and (b) Heart Rate Variability of Light and Moderate Activity}
    \label{fig_7}
\end{figure}

\textit{HR, HRV, and HF-HRV:} The comparison of the mean heart rate over the two activities (light and moderate) was conducted using a paired t-test to assess the potential differences between the two work scenarios. The paired t-test revealed a p-value of 0.0729 with a confidence level of 90 percent, suggesting a significant difference between these two activities. Participants' mean HR (beat per minute) when performing the light and moderate activities were 100.35 and 104.29 bpm, respectively. Figure \ref{fig_7}(a) shows the distribution of the mean HR. The lower mean heart rate in light activity can be associated with decreased distractions due to external stimuli. Since the leaf blower is heavier and noisier than the tablet, the workers could not pay attention to the environment.
Similarly, we compared the heart rate variability over the two activities using a paired t-test to assess the potential differences between the two activities. The paired t-test revealed a p-value of 0.1070, suggesting a confidence level of approximately 90 percent. The mean HRV of the light and moderate activity level were 36.06 and 38.22 bpm, respectively. Figure \ref{fig_7}(b) shows the distribution of the HRV. A decrease in heart rate variability (HRV) indicates a reduction in the adaptability and responsiveness of the autonomic nervous system. This can be associated with an increase in stress when participants performed moderate activity.

\begin{figure}[!htb]
    \centering
    \includegraphics[width=0.48\textwidth]{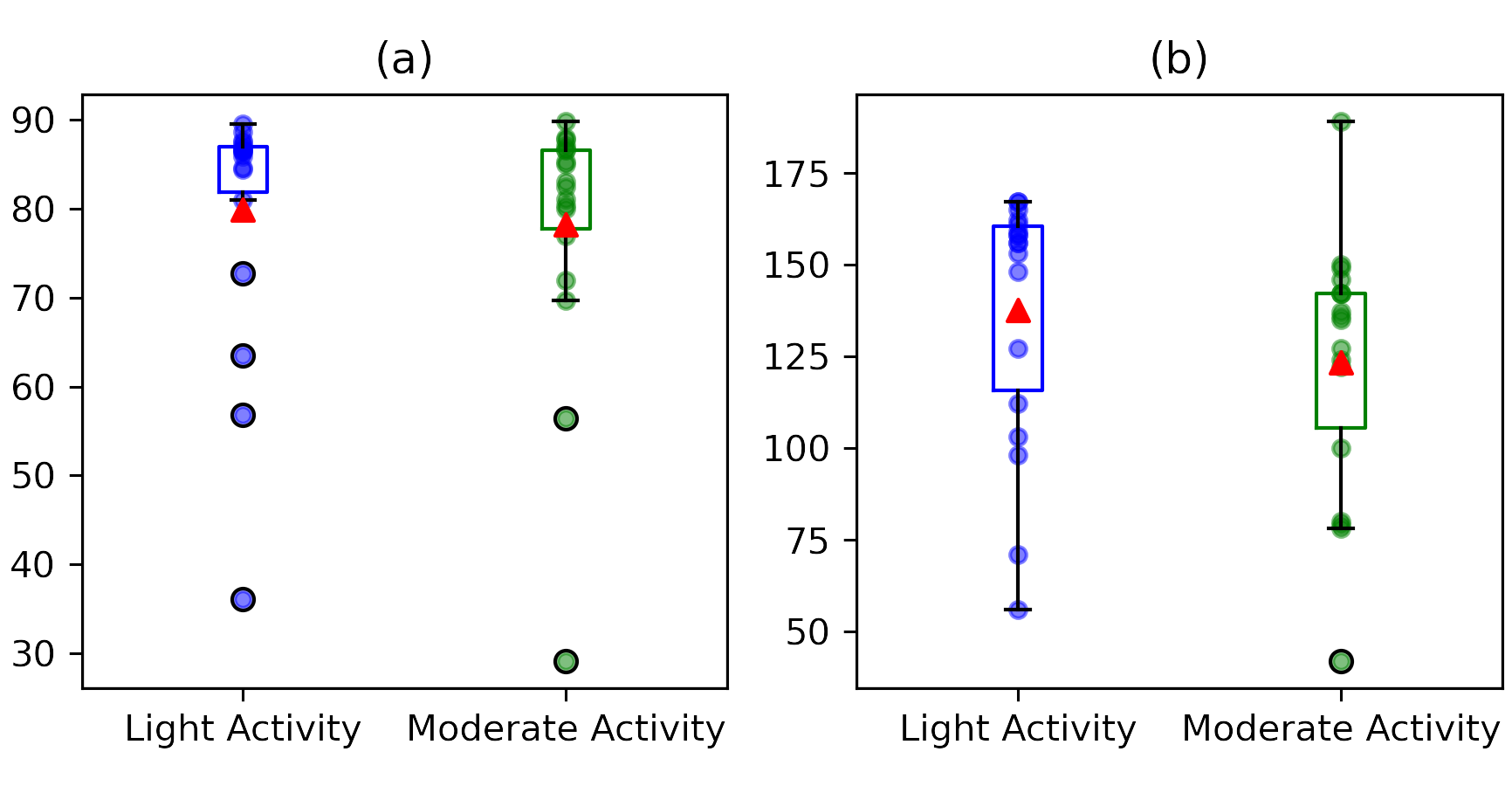}
    \caption{(a) pNN50-HRV and (b) NN50-HRV of Moderate and Light Activity}
    \label{fig_8}
\end{figure}

\begin{figure}[!htb]
    \centering
    \includegraphics[width=0.48\textwidth]{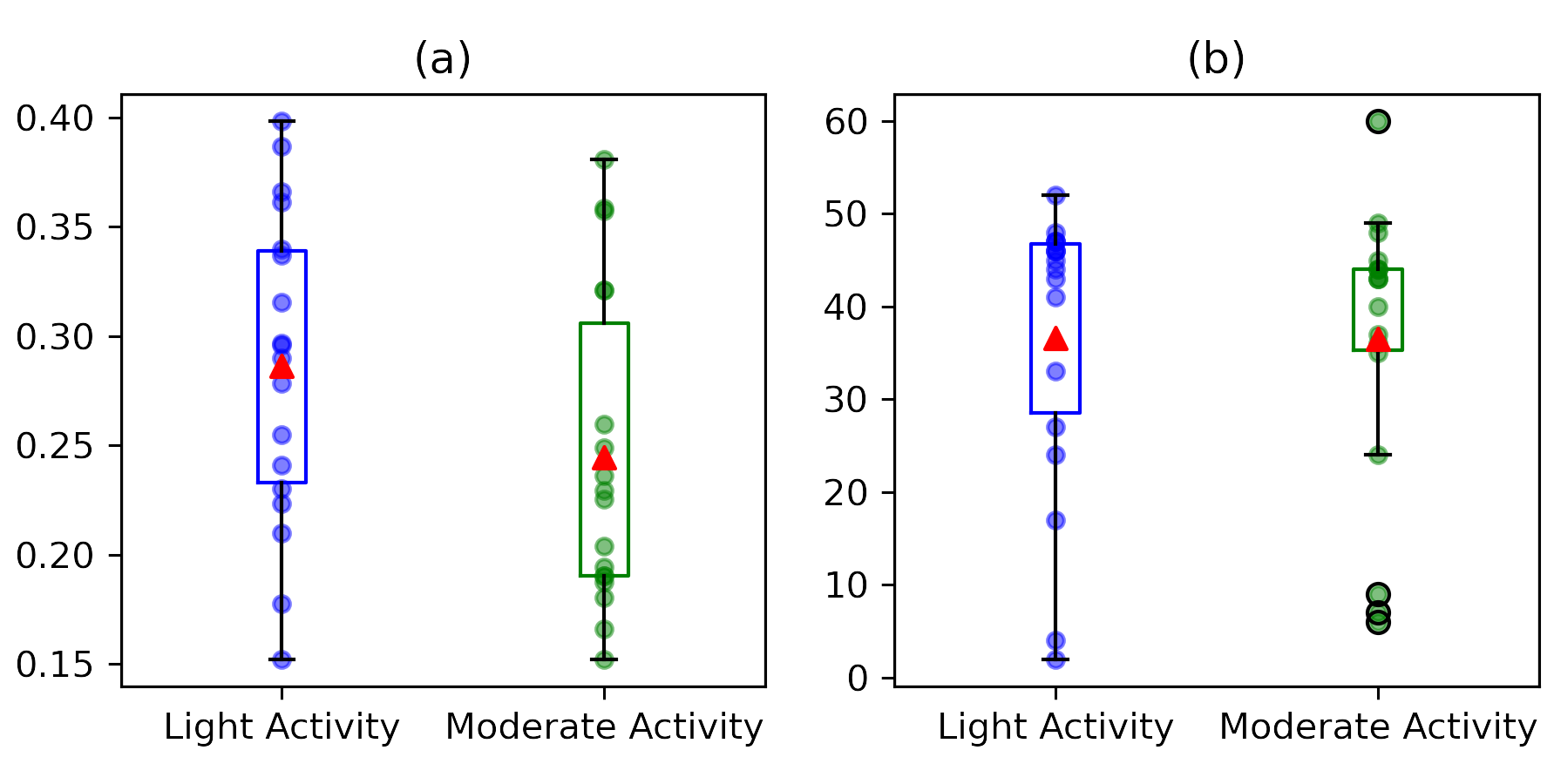}
    \caption{(a) HF-HRV and (b) Number SCR Peaks in Light and Moderate Activity}
    \label{fig_9}
\end{figure}

Another interesting finding from HRV is in the pNN50 and NN50 features. The paired t-test revealed a p-value of 0.7349 and 0.0257, respectively. The NN50 with a confidence level of 95 percent, suggesting a significant difference between these two activities. The moderate activity recorded a lower mean NN50 compared to the light activity. Figure \ref{fig_8}(a) and (b) shows the distribution of the HRV. Both pNN50 and NN50 are used in HRV analysis to assess the balance between the sympathetic and parasympathetic (PNS) branches of the autonomic nervous system \cite{shaffer2017overview}. Although in a low-intensity scenario, the pNN50 has no significant difference as compared to the moderate intensity, lower pnni50 shows lower PNS activity during moderate activity, revealing a higher workload level \cite{kim2018stress}.
Comparing the HF-HRV across the light and moderate activity, the paired t-test shows a significant difference between them with a confidence interval of approximately 90 percent (P-value = 0.1025). The distribution of the HF-HRV is depicted in the Figure \ref{fig_9}(a). This parameter is also associated with the parasympathetic nervous system. Lower HF-HRV values are generally associated with more stress \cite{kim2018stress} during moderate activity.

It should be noted that in this study, we used the  Normal-to-Normal Intervals (NNI) and Inter beat Intervals (IBI) for the calculation of the HRV, which is a measure of the variation in time between each heartbeat. However, the intensity of physical workloads can also affect stress indicators, and distinguishing between the effects of physical workloads and mental stress on these indicators presents a challenge due to their overlapping impacts. When the body is subjected to varying levels of physical demand, the HR increases, but it does not necessarily mean that the body is under mental stress. A higher heart rate, generally resulting from physical activity or stress, is associated with lower HRV. This is because a faster heart rate reduces the time between individual heartbeats, leaving less room for variability. In this study, none of the workload scenarios involved intense physical activities. The results were presented with the assumption that the variations between low and medium physical activities will not have a significant impact on stress indicators. This assumption was based on the understanding that the influence of physical activity on heart rate (HR) and heart rate variability (HRV) becomes more pronounced under conditions of intense physical exertion.

\textit{EDA:} The mean of SCR peaks associated with the cognitive load of participants for the low-intensity and moderate-intensity scenarios were 36.61 (SD=15.43) and 36.5 (SD=15.22), respectively. Also, the paired t-test showed a p-value of 0.9778, suggesting that in a low-intensity scenario, the mean of SCR peaks has no significant difference as compared to the moderate intensity, as shown in Figure \ref{fig_9}(b).

The findings revealed significant physiological differences between light and moderate activities, particularly in HRV measures of HF, RMSSD, NN50, and pNN50, which are used for assessing the autonomic nervous system's balance. The moderate activity demonstrated a reduction in HRV and a lower mean NN50, indicating increased stress levels and a potential decrease in parasympathetic activity. Furthermore, the analysis did not find significant differences in the low and moderate scenarios as measured by the number of SCR peaks, suggesting that the type of physical activity may not impact cognitive stress indicators in controlled conditions.

\section{Conclusion}
This study aimed to assess the stress levels experienced by construction workers as they received warnings during regular roadway work zone tasks, examining the effects within the categories of light or moderate-intensity tasks. The experiment model used physiological signals collected from a wristband-type sensing technology device while they carried out routine highway maintenance work in a VR environment.
Results from 18 participants in the simulated work zone through a virtual environment indicate that (1) workers had a significantly lower heart rate variability and mean heart rate when they did medium-intensity activity than low-intensity activity; (2) moderate activity recorded a lower mean NN50 and pNN50 as compared to the light activity; (3) the HF-HRV across these two activities show a significant difference between them; (4) there are no significant differences between the SCR peaks for the low and moderate activity scenarios. The human-sensing stress recognition model introduced in this study offers two significant contributions: (1) It incorporates workers' physiological characteristics to assess their stress levels while receiving an AR-enabled warning through a combination of audio, visual, and haptic signals. (2) It further clarifies the relationship between worker stress, workload, and activities that demand physical effort.

This study acknowledges a few limitations that should be addressed in future research.  Expanding the sample size of participants to include a more diverse and larger population would enhance the generalizability of the findings. Moreover, the study did not explore the effects of gender, race, age, and disability, areas which future studies could consider to deepen the understanding of stress recognition in varying demographics. Additionally, the use of head-mounted VR displays may introduce motion sickness and dizziness, potentially confounding stress measurements. The varied backgrounds of participants, including their AR/VR experience and onsite experience, could also influence the results, suggesting a need for subgroup analysis to reveal these effects. Furthermore, the experiment was limited to tasks categorized as light and medium in terms of activity level. Future research could benefit from incorporating a broader range of tasks, including those classified as heavy, to more thoroughly assess the impact of workload and physical demand on stress levels.

\bibliographystyle{IEEEtran}
\bibliography{ISARC}  

\end{document}